# p-Type Sb-doped ZnO Thin Films Prepared with Filtered Vacuum Arc Deposition.

T. David, S. Goldsmith, R.L. Boxman
Electrical Discharge and Plasma Laboratory,
Tel Aviv University, Tel Aviv, ISRAEL

## ABSTRACT

Thin p-type Sb-doped ZnO films were grown by filtered vacuum arc deposition (FVAD), on untreated glass samples. The arc cathode was prepared by dissolving Sb into molten Zn. The deposition was performed with 200 A arc current, running for 120-240 s in 0.426 Pa oxygen pressure. The film thickness was 330 – 500 nm. The atomic concentration of Sb in the films was ~1.5%, whereas the O/Zn atomic concentration ratio was ~0.7. Sb incorporation into the polycrystalline ZnO matrix was concluded from XRD analysis. The film resistivity (0.15-0.3 $\Omega$m), carrier density (~$10^{22}$ m$^{-3}$), and carrier type, were determined by Hall effect measurement. The carrier mobility in the p-type films was in the range $(10\text{-}20)\cdot 10^{-4}$ m$^2$/(V·s). The energy gap of doped films, determined from optical transmittance measurements, was approximately 3.39 eV.

## INTRODUCTION

Zinc oxide (ZnO) is a II-VI transparent conducting oxide (TCO), with a wide band-gap of ~3.3 eV at room temperature. In its non-stoichiometric form, it is an n-type semiconductor, and could be grown with resistivity as small as ~$10^{-6}$ $\Omega$·m, i.e., a resistivity only 1-2 orders of magnitude larger than that of metals. These features make ZnO a potential candidate material for use in optoelectronic devices, such as blue LED's and laser diodes (LD's). Thin transparent and electrically conducting ZnO films were deposited using sputtering techniques [1-5], chemical vapor deposition (CVD) [6-8], molecular beam epitaxy (MBE) [9,10], filtered vacuum arc deposition (FVAD) [11-16], and more [17,18].

In contrast to n-type ZnO, p-type ZnO films can be produced only by doping. The various methods used in growing p-type doped ZnO include standard doping, usually with nitrogen [10,14,19-22], co-doping [23,24], or diffusion doping from the substrate material or an interface layer of the dopant, upon using an excimer laser [25-27]. Generally, the grown p-type films have resistivities of $(1\text{-}400)\cdot 10^{-3}$ $\Omega$m, mobility of ~$(0.1\text{-}50)\cdot 10^{-4}$ m$^2$/(V·s), and hole concentration of ~$10^{21}$-$10^{27}$ m$^{-3}$. Aoki et al. reported on p-type ZnO thin films obtained by Sb-doping, using the excimer laser diffusion method [28]. The ZnO layers were epitaxially grown on Si substrates (thickness ~50 nm), and p-type films with resistivity of $8\cdot 10^{-5}$ $\Omega$m, mobility of $1.5\cdot 10^{-4}$ m$^2$/(V·s), and hole concentration of $5\cdot 10^{26}$ m$^{-3}$ were reported, using gold electrodes as contacts. No data on the composition or microstructure of the doped films were reported.

In this paper, we report on the physical characteristics of p-type Sb-doped ZnO thin films prepared by FVAD. Film structure, composition, electrical properties, and optical properties of undoped and doped films are compared and discussed.

## EXPERIMENTAL SYSTEM AND PROCEDURE

The deposition system was described previously in detail [29-31]. It is based on a plasma gun, and a quarter-torus magnetic macroparticle filter, attached to a deposition chamber. Oxygen is introduced near the vicinity of the substrate, and the process is pressure controlled. The cathode is a water-cooled 91 mm diameter Cu cup, filled with a mixture of 99.9% zinc and 3 at. % antimony. A clean-up arc in vacuum was performed to remove any residual surface contamination, before depositing films. The films were produced at a pressure of 0.426 Pa and with an arc current of 200 A. Deposition times were 60 s for undoped films and 120, and 240 s for doped films. In each run, three samples, 1x1x0.1 cm glass slides, were placed inside the central region of the plasma beam, obtaining the same thickness with uniformity of ~10% [29].

The substrates were not biased, and were water-cooled through the substrate holder throughout the deposition process. The substrate temperature was not monitored in the process, but could be estimated to be not higher than several tens of degrees [32].

Sample thickness was measured with an alpha-step profilometer. X-ray photoelectron spectroscopy (XPS) was used to determine film composition, using a PHI scanning 5600 AES/XPS multi-technique system. Depth information was obtained by sputtering with Ar$^+$ ions through the film, combined with AES/XPS analysis. X-ray diffraction (XRD) studies were performed using a Scintag X-ray diffractometer equipped with a Cu anode (Cu K$\alpha$, $\lambda$=0.1541 nm). The diffraction pattern intensity, location, and width were determined with commercial software [33]. The electrical properties were measured with a HEM 2000 Hall-effect measurement system [34], utilizing the Van der Pauw method. The substrates were bonded to the system's substrate holder contacts using a silver paste. Optical transmission was measured by a Minuteman 305MV monochromator equipped with a calibrated photo-multiplier, and a continuous Hg light source.

## RESULTS AND DISCUSSION

The electrical resistance of doped samples obtained with 60 s deposition time was larger than 2000 kΩ. Their XRD spectra were very weak, not showing any diffraction lines. In contrast, undoped ZnO films obtained with the same deposition system and parameters had a polycrystalline structure [29].
The deposition rate obtained with the doped cathode was lower than obtained with a pure Zn cathode. The thickness of the films deposited for 120, and 240 s (hereafter labeled "120 s" and "240 s", respectively) was 330, and 500 nm, respectively, while the undoped sample deposited for 60 s with the same pressure and arc current as the doped samples was 430 nm thick.

### Film Composition

The composition of the undoped, 120 s, and 240 s doped samples is shown in table 1, with the accuracy estimate for the XPS diagnostics. The atomic concentration of Sb in the films was lower than the cathode doping percentage. This could be attributed to having a non-uniform mixture of Zn-Sb in the cathode. The oxygen-to-zinc ratios listed in table 1 were similar to those previously determined in undoped films deposited with the same FVAD system [29,32]. Apart from these elements, there was some presence of Cu, C, and N in the bulk of the films, all less than 1 at. %. The Cu and C contaminations were seen and addressed before [29], and the N contamination could be attributed to residues of nitrogen introduced into the system before opening it to the atmosphere at the end of the deposition procedure. Evidently, there was some diffusion of N into the films. The N concentrations were merely 0.13, and 0.26 at. %, in the 120 s, and 240 s doped films, respectively. No nitrogen was found in undoped films. As N is also a common p-type dopant, it also could be contributing to the electrical properties of the films presented below. However, as the ionization potential of Sb is much lower than that of N, and as the Sb atomic concentration is one order of magnitude larger than that of N, it is reasonable to assume that the principal donor should be the Sb.

### Film Microstructure

The XRD spectra obtained for the undoped, 120 s and 240 s doped samples are shown in figure 1 together with the XRD lines of undoped samples. For clarity, the doped film's spectra were shifted upward relative to that of the undoped one.

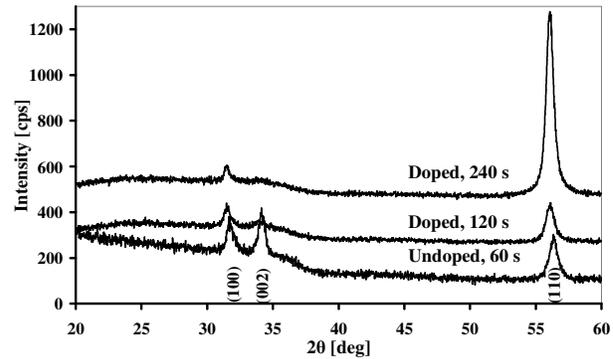

**Figure 1.** X-ray diffraction spectra for of the undoped and doped samples. For clarity, the spectra of the doped samples deposited for 120 and 240 s were shifted upwards by 200 and 400 cps, respectively.

The spectra show well-defined XRD lines, indicating a polycrystalline structure. The grains had preferred orientations, which were dependent on the doping. For the undoped film, the preferred orientation was parallel to the c-axis (the (002) line indicating hexagonal-wurtzite structure), whereas the orientation of the crystallographic planes of the doped films was less well-defined, showing also diffraction by the planes parallel to the (110) orientation. The intensities of the (002) and (100) diffraction lines were substantially lower in the doped films in comparison with undoped films, and the (110) line dominated. The grain sizes, determined from the (110) diffraction line according to Scherer's relation:

$$D = \frac{0.94\lambda}{\omega \cdot Cos(\theta)}$$

were 11, 11, and 13 nm for the undoped, 120 s doped, and 240 s doped samples, respectively. In this relation D is the grain size in nm, λ is the diffractometer wavelength in nm, ω is the (110) line full width at half-maximum after subtracting instrumental width, and θ is the diffraction angle. The diffraction lines were also shifted to lower angles in the doped samples. The (110) line peak position was at 2θ=56.35° in the undoped sample, and was at 56.09°, and 56.07° in the 120 s and 240 s doped samples, respectively. This shift in line position towards smaller angles indicates an increase in lattice constant magnitude between the doped and undoped samples [35].
The effect of the deposition time or thickness on the film structure, seen in the doped film spectra, is similar to that seen in undoped films. The change both in preferred orientation and line position found in the doped samples relative to the undoped samples, as seen in the XRD analysis, indicated that Sb atoms were incorporated into the ZnO matrix.

**Table 1. Composition and electrical properties of undoped and doped films**

| Cathode Doping [at. %] | Sb at bulk [at. %] | O/Zn Ratio | Deposition time [s] | Thickness [nm] | ρ [Ω·m] | M [m²/(V·s)] | Carrier Density [m⁻³] | Type |
|---|---|---|---|---|---|---|---|---|
| 0 | 0 | 0.68±0.03 | 60 | 430 | $0.11 \cdot 10^{-3}$ | $11.6 \cdot 10^{-4}$ | $3.77 \cdot 10^{25}$ | n |
| 3 | 1.5±0.3 | 0.721±0.03 | 120 | 330 | $156 \cdot 10^{-3}$ | $8.56 \cdot 10^{-4}$ | $4.65 \cdot 10^{22}$ | p |
| 3 | 1.45±0.3 | 0.724±0.03 | 240 | 500 | $283 \cdot 10^{-3}$ | $21.45 \cdot 10^{-4}$ | $1.03 \cdot 10^{22}$ | p |

**Electrical Properties**

The electrical properties of the undoped and doped films are shown in table 1. The undoped film showed n-type behavior, while the doped films showed p-type behavior. It could be seen that the doped films had higher resistivity than that of the undoped sample. The charge carrier density was also much lower in the doped films than in the undoped ones, hence the mobility rates were generally comparable between the undoped and doped samples. Stoichiometric ZnO is an insulator, and when there is a deviation from stoichiometry, excessive Zn, it is an n-type semiconductor whose conductivity is attributed to the excessive Zn acting as donor [1]. Sb is a V-element, thus it has the potential to be an acceptor in Sb:ZnO. Thus, Sb:ZnO is doped both by donors and by acceptors. In the present case, the excess of Zn and dopant Sb 'compete' in determining the electrical properties of the films. The data shown here indicate that the doping of ~1.5 at. % Sb resulted in a p-type semiconducting film in spite of about 5% excess in Zn concentration.

**Optical Properties**

The optical properties of the doped films were determined by studying the optical transmission. The wavelength band pass was set to 0.8 nm, and the transmission was measured in the wavelength region 320 to 600 nm. The real and imaginary components of the film's refraction index were obtained by least squares fitting of an expression for the transmission spectrum for Sb:ZnO films to the measured data. The expression was derived by considering light absorption in the Sb:ZnO film, as well as reflection and transmission at three boundaries: (1) air-film, (2) film-glass, and (3) glass-air [29]. The values of k were in the range 0.18 ($\lambda$ =310 nm) to 0.08 ($\lambda$ =460 nm). The value at $\lambda$ =460 nm was much smaller than that of undoped ZnO films deposited by the same system with the same deposition parameters [29]. The values of the real component n decreased monotonically from 2.39 at 340 nm to 2.08 at 440 nm, which is an asymptotic value. Unlike the k values, the n values agree with those determined for undoped films. This film's thickness was also determined from this fit, and was approximately 100 nm.

The energy gap, Eg, of the film was derived from a linear extrapolation of the plot of the expression $(-\ln(1/T) \cdot h\nu)^2$ in the range $\lambda$= (310 –390) nm against $h\nu$ (figure 2), where T is the film transmission at wavelength $\lambda$, $\nu$ is the radiation frequency at the corresponding wavelength, and h is Planck's constant. The value of Eg is approximately 3.39 eV. However, the optical band gap for pure bulk ZnO is known to be 3.2 eV [30]. This difference can be interpreted as the Burnstein-Moss effect [30], which is supported by the XPS and XRD data.

The optical and the electrical data agree with regard to the values of the resistivity and the index k. Carrier density is lower in the doped film compared to undoped ZnO, and the resistivity is higher by two orders of magnitude than that of the undoped film [29]. The observation of higher T, or lower k index, for the doped films properly correlates with higher electrical resistivity, provided the latter is caused by lower carrier density.

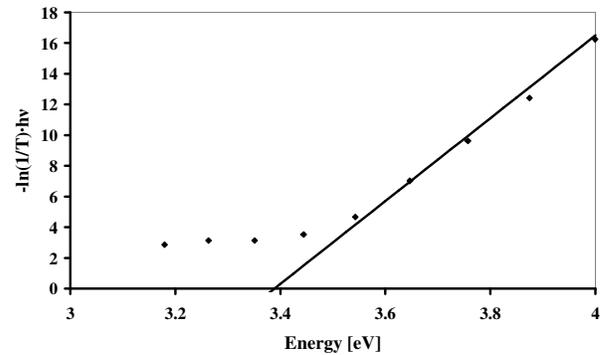

**Figure 2.** $(-\ln(1/T) \cdot h\nu)^2$ as a function of wavelength $\lambda$, near the absorption edge. The solid line is the linear extrapolation used to determine the energy gap Eg.

**CONCLUSION**

In conclusion, we demonstrated the deposition of p-type ZnO thin films by FVAD with Sb-doping. The electrical properties of the films, resistivity (0.15-0.3) Ω·m, mobility ~(10-20)·10⁻⁴ m²/(V·s), and hole concentration of ~10²² m⁻³, are comparable with the properties of the few reports available on p-type ZnO. The energy band gap Eg was determined from optical analysis to be ~3.39 eV. The real part of the refractive index n decreased monotonically as a function of wavelength from 2.39 to an asymptotic value of 2.08 at $\lambda$=440nm, and the imaginary part of the refractive index k decreased from 0.18 to 0.08 for λ=310 to 460 nm, respectively. Furthermore, the present study suggests that FVAD is a suitable method for p-type ZnO thin film deposition.

The authors thank Mr. Y. Zidon, Dr. L. Burstein and Dr. Yu Rosenberg for the Hall effect, XPS, and XRD measurements. This research was partially supported by a grant from the Gordon Center for Energy Studies at Tel Aviv University.

**REFERENCES**

1. K. Ellmer, J. Phys. D: Appl. Phys. 33, R17-R32 (2000), gives a comprehensive review of studies on sputtered ZnO.

2. K. Ellmer, J. Phys. D: Appl. Phys. 34, 3097-3108 (2001).

3. E. M. Bachari, G. Baud, S. Ben Amor, M. Jacquet, Thin Solid Films 348, 165-172 (1999).

4. K. B. Sundaram and A. Khan, Thin Solid Films 295, 87-91 (1997).

5. L. Meng, C. P. Moreira de Sá, M. P. dos Santos, Appl. Surf. Sci. 78, 57-61 (1994).

6. K. Kaiya, K. Omichi, N. Takahashi, T. Nakamura, S. Okamoto, H. Yamamoto, Thin Solid Films 409, 116 (2002).

7. B. S. Li, Y. C. Liu, D. Z. Shen, Y. M. Lu, J. Y. Zhang, X. G. Kong, X. W. Fan, Z. Z. Zhi, J. Vac. Sci. Technol. A 20(1), 265-269 (2002).

8. X. Li, Y. Yan, T. A. Gessert, C. L. Perkins, D. Young, C. DeHart, M. Young, T. J. Coutts, J. Vac. Sci. Technol. A 21(3), 1342-1346 (2003).

9. H. Kato, M. Sano, K. Miyamoto, Yao T., J. Cryst. Growth 237-239, 538-543 (2002).

10. D. C. Look, D. C. Reynolds, C. W. Litton, R. L. Jones, D. B. Eason, G. Cantwell, Appl. Phys. Lett. 81(10), 1830-1832 (2002).

11. X. L. Xu, S. P. Lau, B. K. Tay, Thin Solid Films 398-399, 244-249 (2001).

12. X. L. Xu, S. P. Lau, J. S. Chen, G. Y. Chen, B. K. Tay, J. Cryst. Growth 223, 201-205 (2001).

13. Y. G. Wang, S. P. Lau, H. W. Lee, S. F. Yu, B. K. Tay, X. H. Zhang, K. Y. Tse, H. H. Hng, J. Appl. Phys. 94(3), 1597-1604 (2003).

14. Y. G. Wang, S. P. Lau, X. H. Zhang, H. W. Lee, H. H. Hng, B. K. Tay, J. Cryst. Growth 252, 265-269 (2003).

15. X. L. Xu, S. P. Lau, J. S. Chen, Z. Sun, B. K. Tay, J. W. Chai, Mat. Sci. Sem. Procss. 4, 617 (2001).

16. H. Takikawa, K. Kimura, R. Miyano, T. Sakakibara, Thin Solid Films 377-378, 74-80 (2000).

17. J. F. Muth, R. M. Kolbas, A. K. Sharma, S. Oktyabrsky, H. Narayan, J. Appl. Phys. 85(11), 7884-7887 (1999).

18. M. J. Alam and D. C. Cameron, J. Vac. Sci. Technol. A 19(4), 1642-1646 (2001).

19. X. Li, Y. Yan, T. A. Gessert, C. L. Perkins, D. Young, C. DeHart, M. Young, T. J. Coutts, J. Vac. Sci. Technol. A 21(4), 1342-1346 (2003).

20. K. Minegishi, Y. Koiwai, Y. Kikuchi, K. Yano, M. Kasuga, A. Shimizu, Jpn. J. Appl. Phys. 36, L1453-L1455 (1997).

21. X. –L. Gou, H. Tabata, T. Kawai, J. Cryst. Growth 223, 135-139 (2001).

22. N. Y. Garces, N. C. Giles, L. E. Halliburton, G. Cantwell, D. B. Eason, D. C. Reynolds, D. C. Look, Appl. Phys. Lett. 80(8), 1334-1336 (2002).

23. M. Joseph, H. Tabata, H. Saeki, K. Ueda, T. Kawai, Physica B 302-303, 140-148 (2001).

24. T. Yamamoto, Phys. Stat. Sol. (a) 193 (3), 423-433 (2002).

25. T. Aoki, Y. Hatanaka, D. C. Look, Appl. Phys. Lett. 76(22), 3257-3258 (2000).

26. Y. Hatanaka, M. Niraula, A. Nakamura, T. Aoki, Appl. Surf. Sci. 175-176, 462-467 (2001).

27. Y. R. Ryu, S. Zu, D. C. Look, J. M. Wrobel, H. M. Jeong, H. W. White, J. Cryst. Growth 216, 330-334 (2000).

28. T. Aoki, Y. Shimizu, A. Miyake, A. Nakamura, Y. Nakanishi, Y. Hatanaka, Phys. Stat. Sol (b) 229(2), 911-914 (2002).

29. T. David, S. Goldsmith, R. L. Boxman, Thin Solid Films 447-448C, 61-67 (2003).

30. L. Kaplan, V. N. Zhitomirsky, S. Goldsmith, R. L. Boxman, I. Rusman, Surf. Coat. Technol. 76-77, 181-189 (1995).

31. L. Kaplan, A. Ben-Shalom, R. L. Boxman, S. Goldsmith, U. Rosenberg, M. Nathan, Thin Solid Films 253, 1-8 (1994).

32. T. David, S. Goldsmith, R. L. Boxman, "Depenedence of ZnO Thin Film Properties on Filtered Vacuum Arc Deposition Parameters", arxiv e-print archive: cond-mat/0501374.

33. KyPlot software, version 2, available from K. Yoshioka (Kyence Inc.), KyPlot homepage: http://www.qualest.co.jp/Download/KyPlot/kyplot_e.htm

34. The HEM 2000 measurement system is by Bridge Technology. Specifications on the system can be found at http://four-point-probes.com/egk.html.

35. S. Maniv, W. D. Westwood, E. Colombini, J. Vac. Sci. Technol. 20, 162-170 (1982)

36. G.H. Lee, Y. Yamamoto, M. Kourogi, M. Ohtsu, Thin Solid Films 386,117 (2000